\documentclass[11pt,epsfig,citesort,svgnames,dvipsnames]{article}
\usepackage{amsfonts,amssymb,amsmath,amsthm,amscd,latexsym}
\usepackage[latin1]{inputenc}
\usepackage{graphicx}
\usepackage[all]{xy}
\usepackage{caption}
\usepackage{subcaption}
\usepackage{lmodern}
\usepackage{microtype}
\usepackage{verbatim}
\usepackage{indentfirst}
\usepackage{marginnote} % for \marginnote, a replacement for \marginpar that works everywhere
\usepackage{ifthen} % for \ifthenelse
\usepackage{tensor}
\usepackage{tikz}
\usetikzlibrary{matrix,arrows}
\usepackage[official]{eurosym}
\usepackage[numbers,sort&compress]{natbib}
\usepackage{xcolor}
\usepackage{hyperref}
\hypersetup{
colorlinks,
citecolor=WildStrawberry,
linkcolor=Blue, 
urlcolor=Green}

\def\be{\begin{equation}}
\def\ee{\end{equation}}
\def\bea{\begin{eqnarray}}
\def\eea{\end{eqnarray}}

\newcommand{\sect}[1]{\setcounter{equation}{0}\section{#1}}

 \def\1{\'{\i}}

\newcommand{\bq}{\mathbf{q}}
\newcommand{\bp}{\mathbf{p}}

\def\dd{{\rm d}}
\newcommand\Om\Omega

\def\la{{\lambda}}
\newcommand{\e}{{\mathrm e}}

\begin{document}

\

 \
 \smallskip
 
\vskip1cm

\begin{center}
 {\LARGE{\bf{On  Hamiltonians with position-dependent mass
 \\[6pt] from Kaluza-Klein compactifications}}}

\bigskip
   
\bigskip

\begin{center}
{\sc \'Angel Ballesteros, Iv\'an Guti\'errez-Sagredo, Pedro Naranjo}
\end{center}

\noindent
Departamento de F\1sica,  Universidad de Burgos,
09001 Burgos, Spain\\ ~~E-mail: angelb@ubu.es,  igsagredo@ubu.es,  pnaranjo@ubu.es\\

\end{center}

\medskip
\medskip
\medskip

\begin{abstract}
In a recent paper (J.R. Morris, Quant. Stud. Math. Found. 2 (2015) 359), an inhomogeneous compactification of the extra dimension of a five-dimensional Kaluza-Klein metric has been shown to generate a position-dependent mass (PDM) in the corresponding four-dimensional system. As an application of this dimensional reduction mechanism, a specific static dilatonic scalar field has been connected with a PDM Lagrangian describing a well-known nonlinear PDM oscillator. Here we present more instances of this construction that lead to PDM systems with radial symmetry, and the properties of their corresponding inhomogeneous extra dimensions are compared with the ones in the nonlinear oscillator model. Moreover, it is also shown how the compactification introduced in this type of models can alternatively be interpreted as a novel mechanism for the dynamical generation of curvature.
\end{abstract}

\bigskip\bigskip 

  
 \medskip


%\newpage

%%%%%%%%%%%%%%%%%%%%%%%%%
\sect{Introduction}

The classical Hamiltonian systems of the type 
\be
{\cal H}(\bq,\bp)={\cal T}(\bq,\bp)+{\cal U}(\bq)=
\frac{\bp^2}{2\, {\cal M}(\bq)}+{\cal U}(\bq) ,
\label{uno}
\ee
where $(\bq,\bp)$ are conjugate coordinates and momenta with canonical Poisson bracket $\{q_i,p_j\}=\delta_{ij}$,  describe the motion on the ND Euclidean space of a PDM particle with mass function ${\cal M}(\bq)$  under the potential ${\cal U}(\bq)$ (see, for instance,~\cite{Morris, Roos, Young, mass2, mass3, mass4, Gritsev, Quesnea, Quesne, Koc, Quesnec, Schd, Mustafa, mass5, SAG, MR, Levai, Chargui} and references therein). In a recent paper~\cite{Morris}, these systems have been shown to arise from the compactification of an inhomogeneous extra dimension in a Kaluza-Klein framework, in such a way that the variable mass function ${\cal M}(\bq)$ turns out to be the Euclidean footprint of the inhomogeneity in the size of the extra dimension. 

As a guiding example, the two-dimensional nonlinear oscillator Hamiltonian with mass function and potential given by
\be
{\cal M}(\bq)=\frac{m_0}{1+\la \bq^2},
\qquad
{\cal U}(\bq)=\frac{m_0\,\alpha^2\,\bq^2}{2(1+\la \bq^2)},
\label{noh}
\ee
where $m_0,\alpha$ and $\la$ are real parameters,
was shown in~\cite{Morris} to arise from a dilatonic coupling function emerging from the compactification of a given extra dimension. This Hamiltonian belongs to a well-known family of systems, since its one-dimensional analogue is just the (superintegrable) nonlinear oscillator introduced by Mathews and Lakhsmanan~\cite{ML}, and its two and three-dimensional generalisations are related to the isotropic oscillator on the sphere (if $\la>0$) or on the hyperbolic space (if $\la<0$) (see~\cite{Higgs, CRS04, CRS07, CRSjmp, monopoles} for details). The exact solvability of this kind of nonlinear superintegrable PDM systems in both classical and quantum contexts, together with their applicability in different physical situations, has motivated a number of recent investigations on these models (see~\cite{darbouxPLA, Gandarias, darboux, CRSjpa, SMjpa, CRSjmp12, SM, SMepj, QuesnePLA, taubNUT, QuesneEPL} and references therein). 

As will be shown in the sequel, the viewpoint presented in~\cite{Morris} opens up the path to the consideration of other PDM Hamiltonians and to analyse their associated Kaluza-Klein models. The main motivation to consider these new examples consists in the fact that  they can be thought of as suitable toy-models in order to explore the classical and quantum consequences of the compactification of different classes of inhomogeneous extra dimensions, as well as the dynamical and geometric properties arising from it. In particular, we will present two new infinite families of models that generalise the results presented in~\cite{Morris} and whose classical and quantum integrability properties seem worth to be studied.

The structure of the paper is as follows. In the next section, the Morris construction of PDM systems as a result of a compactification procedure will be sketched. Section 3 will be devoted to the study of two new families of two-dimensional PDM models with radial symmetry, and to compare the results so obtained with the ones coming from the nonlinear oscillator Hamiltonian proposed in~\eqref{noh}. 
Moreover, in the last section we propose an alternative geometric interpretation of all the previous results  based on the fact that PDM Hamiltonians can also be interpreted as motions of a particle with constant mass on spaces with (nonconstant) curvature, and this idea is tested on the families of solutions that have been previously introduced. Therefore, it could be thought of the inhomogeneous compactifications here described as giving rise to effective motions on curved spaces, where the local curvature of the space is related to the size of the extra dimension. Finally, a concluding section closes the paper.

%%%%%%%%%%%%%%%%%%%%%%%%%%%%%%%%%%%%%%%%%%%%%%%%%%%%%%%%%

\sect{PDM systems from compactification}

In this Section we will summarise the interesting connection proposed in~\cite{Morris}, that links dimensional reduction from a 5D Kaluza-Klein metric with effective PDM systems (we will concentrate on the essential results; for further details, see~\cite{Morris} and references therein). Such a construction starts from a metric on a 5D spacetime of the form
\be
ds^2=g_{\mu \nu}(x^\alpha) \, dx^\mu\,dx^\nu- [b(x^\alpha)]^2\,dy^2,
\ee
where the (3+1) metric $g_{\mu \nu}$ has signature $(+,-,-,-)$ 
and the last coordinate $y$ corresponds to the additional dimension, to be later compactified. The 5D theory is thus defined by introducing an action that includes gravitation and matter, and the 4D theory arises through dimensional reduction by performing an integration over the extra dimension. In such a 4D theory, the function $b(x^\alpha)$ behaves as a scalar field that is nonminimally coupled to the Ricci scalar. Finally, by performing a conformal transformation of the 4D metric, the theory can be rewritten in the Einstein frame, where the scalar field $b(x^\alpha)$ decouples from the curvature but gets coupled to the matter sector. More explicitly, the dilaton field $\phi$ turns out to be related to the extra dimension scale function $b$ through
\be
\phi(x^\alpha)=\frac{1}{a}\,\ln b(x^\alpha),
\ee
where $a$ is a constant. Moreover, by writing the matter action in the Einstein frame, the effective mass of the  particle is found to be
\be
m(x^\alpha)=m_0\,[b(x^\alpha)]^{-1/2},
\ee
which implies that the  compactification (which is inhomogeneous, since the size $b(x^\alpha)$ of the extra dimension depends on the spacetime coordinates) gives rise to an effective model on the non-enlarged 4D spacetime in which the particle has a PDM given by $m(x^\alpha)$. 

After performing the appropriate  nonrelativistic limit of the interaction term, the full Galilean Lagrangian for the particle is
\be
L=\frac{1}{2}\,m(x^\alpha)\,[\delta_{ij}\,u^i\, u^j - U_0(x^\alpha)],
\ee
where $u^i$ are the components of the velocities and $U_0(x^\alpha)$ is the potential that describes the interaction of the particle with non-gravitational forces. Obviously, if we consider a static field $\phi(x^1\equiv q_1,x^2\equiv q_2,x^3\equiv q_3)$, the Hamiltonian version of this Lagrangian is 
\be
{\cal H}(\bq,\bp)= \frac{\bp^2}{2\, m(\bq)} + \frac{
m(\bq)}{2}\,U_0(\bq)=
 \frac{\bp^2}{2\, m_0\,[b(\bq)]^{-1/2}} + \frac{
m_0\,[b(\bq)]^{-1/2}}{2}\,U_0(\bq),
\ee
which means that
\be
{\cal M}(\bq)=m_0\,[b(\bq)]^{-1/2},
\qquad
{\cal U}(\bq)=\frac{
m_0\,[b(\bq)]^{-1/2}}{2}\,U_0(\bq)=\frac{
{\cal M}(\bq)}{2}\,U_0(\bq).
\label{potential}
\ee

In particular, the paper~\cite{Morris} considers a static and radially symmetric dilaton field background in which the field $\phi$ has to satisfy the Liouville equation
\be
\nabla^2\phi=-\frac{2\,\Lambda}{3\,a}\e^{-a\phi},
\label{dilaton}
\ee
where $\Lambda$ plays the role of the cosmological constant and $\nabla^2 \equiv \partial_x ^2 + \partial_y ^2$ stands for the two dimensional Laplacian. If we consider the function
$
b=e^{a\,\phi}
$, 
the equation for $b^{-1}$ reads
\be
\nabla^2(\ln b^{-1})=\frac{2\,\Lambda}{3}\,b^{-1}.
\label{dilatonb}
\ee
As a consequence,  the mass function $b^{-1/2}$ has to be a solution of the equation
\be
\nabla^2(\ln b^{-1/2})=\frac{\Lambda}{3}\,[b^{-1/2}]^2.
\label{dilatonm}
\ee

An interesting geometric interpretation of the Liouville equation~\eqref{dilatonb} arises if we consider a two-dimensional metric 
\be
\dd s^2=g_{11}(q_1,q_2)\,\dd 
q_1^2+g_{22}(q_1,q_2)\,\dd  q_2^2,
\ee
whose Gaussian curvature $K$ can directly be computed through~\cite{Berry}
\be
K=\frac{-1}{\sqrt{g_{11} g_{22}}}\left\{ \frac{\partial}{\partial
q_1} \left( \frac{1}{\sqrt{g_{11}}} \frac{\partial
\sqrt{g_{22}}}{\partial q_1}
\right)+
\frac{\partial}{\partial
q_2} \left( \frac{1}{\sqrt{g_{22}}} \frac{\partial
\sqrt{g_{11}}}{\partial q_2}
\right)\right\} .
\label{ccc}
\ee
Now, if the metric possesses the conformally flat form
\be
\dd s^2=e^{\psi(q_1,q_2)}\,(\dd q_1^2 + \dd q_2^2),
\label{metric}
\ee
it is straightforward to check that the Gaussian curvature $K$ reads
\be
K=-\frac12 e^{-\psi}\,\nabla^2\psi,
\ee
which is just the equation~\eqref{dilatonb}, where
\be
b^{-1}=e^{\psi},\qquad \Lambda=-3 K.
\ee
Therefore, solutions to~\eqref{dilatonb} are provided by the 2D conformally flat  metrics associated with spaces of constant Gaussian curvature $K=-\Lambda/3$.

As was shown in~\cite{Morris}, a   particular solution $\phi(q_1,q_2)$ to the two-dimensional equation~\eqref{dilatonb} is given by
\be
e^{\psi(q_1,q_2)}=b^{-1}(q_1,q_2)=\frac{C^2}{(1+\la\,r^2)^2},
\qquad
r^2=q_1^2+q_2^2,
\label{solK}
\ee
where $C$ is a real constant. Indeed, this solution is consistent with the geometric interpretation stated above, since the conformal factor~\eqref{solK} corresponds to the spaces with constant Gaussian curvature proportional to $\lambda$. These spaces have been studied making use of proyective Poincar\'e coordinates in~\cite{monopoles}.
In this case, the corresponding (two-dimensional) PDM Hamiltonian~\eqref{uno} is characterised by a mass function which is the square root of the conformal factor $e^\psi$, namely 
\be
{\cal M}(\bq)=m_0\,\frac{C}{(1+\la\,\bq^2)},
\qquad
{\cal U}(\bq)=\frac{
m_0\,C}{2\,(1+\la\,\bq^2)}\,U_0(\bq),
\label{higgs}
\ee
where we are forced to take $C>0$ in order to have a positive mass function. 

At this point it is worth stressing that the interaction potential $U_0(\bq)$ is not provided by the compactification approach, and gives an additional freedom for the model. Therefore, it would be useful for further physical explorations to see whether a choice for $U_0(\bq)$ can be made, such that the total effective PDM Hamiltonian be exactly solvable (in both the classical and quantum cases). This is exactly the case for the function ${\cal M}(\bq)$ in~\eqref{higgs} if we take $U_0(\bq)=\alpha^2\,\bq^2$, since then
\be
{\cal U}(\bq)=\frac{
m_0\,C\,\alpha^2\,\bq^2}{2\,(1+\la\,\bq^2)},
\label{Morrispot}
\ee
thereby the exactly solvable nonlinear oscillator Hamiltonian introduced in~\cite{ML,CRS04} arising. Also, the corresponding quantum model can exactly be solved~\cite{CRS07, CRSjmp, CRSjpa, SM}. In this context, if the size of the extra dimension is assumed to be given by $b(r)$ (which is the inverse of the conformal factor) 
\be
b(r)= \frac{1}{C^2}\,(1+\la\,r^2)^2,
\label{Morrisb}
\ee
this means that as long as $r$ grows, the extra dimension becomes larger and the effective mass of the particle becomes small. Therefore, the energies of the bound states of the system could be indeed affected by the existence of an extra compactified dimension.

%%%%%%%%%%%%%%%%%%%%%%%%%%%%%%%%%%%%%%%%%%%%%%%%%%%%%%%%%%%%%%%%%%%     

%%%%%%%%%%%%%%%%%%%%%%%%%%%%%%%%%%%%%%%%%%%%%%%%%%%%%%%%%%%%%%%%%%%     

%%%%%%%%%%%%%%%%%%%%%%%%%%%%%%%%%%%%%%%%%%%%%%%%%%%%%%%%%

\sect{Two new families of models}

Having in mind the previous construction, it seems appropriate to explore other Kaluza-Klein compactifications and to interpret them in the context of PDM Hamiltonians. We follow the same construction, by assuming a static two-dimensional and radially symmetric dilaton field arising from the compactification, where the corresponding scalar field $\phi=\frac{1}{a}\,\ln b$  is thus given by a solution of the equation
\be
\left(\partial_{q_1}^2+\partial_{q_2}^2\right) \ln b^{-1} ({q_1},{q_2})=\frac{2\,\Lambda}{3}\,b^{-1} ({q_1},{q_2}).
\label{2DLiouville}
\ee
It can be proven~\cite{Gibbons} that $b^{-1} ({q_1},{q_2})$ is a solution to this equation if a holomorphic function $f\left(\zeta\right)$ (with $\zeta=q_1+i\,q_2$) can be found such that
\be
b^{-1}(r)=C^2\frac{\left|f'\left(\zeta\right)\right|^2}{\left(1\pm\left|f\left(\zeta\right)\right|^{2}\right)^2},
\qquad
\Lambda=\mp\,\left| \Lambda \right| \,,
\label{diff}
\ee
where $r=\left| \zeta \right|$ and $C$ is a real constant. Henceforth, we will denote the functions and constants that corresponds to each of the two possible signs in~\eqref{diff} by $b_\pm$ and $C_\pm$, respectively.

In particular, the holomorphic function
\be
f\left(\zeta\right)=A\,\zeta,
\ee
where $A \in \mathbb{C}$, gives rise to the conformal factor
\be
e^{\psi(r)}=b_\pm^{-1}(r)=C_\pm^2\frac{{\lambda}}{\left(1\pm\lambda\,r^2\right)^2},
\ee
where $\lambda=\left| A \right|^2$. If we compute the Gaussian curvature $K$  for the corresponding metric~\eqref{metric}, we get in the `$+$' case that
\be
K_+=\frac{4}{C_+^2}=-\frac{\Lambda}{3}>0,
\ee
which implies that $\Lambda<0$, and in the `$-$' case we have
\be
K_-=-\frac{4}{C_-^2}=-\frac{\Lambda}{3}<0,
\ee
which is consistent with taking $\Lambda>0$. 

Therefore, from the viewpoint of the mass functions we encounter two possibilities, provided that  $C_\pm\in \mathbb{R}^+$ in order to have a non-vanishing positive mass:
\begin{itemize}
\item The `$+$' case: it corresponds to $\Lambda<0$. Then, $C_+=\sqrt{12/(-\Lambda)}$ and the solution reads
\be
b_+^{-\frac{1}{2}}(r)=\sqrt{\frac{12}{(-\Lambda)}}\frac{\sqrt{\lambda}}{\left(1+\lambda\,r^2\right)}\,.
\ee
This mass function is just the one for the nonlinear oscillator case~\eqref{Morrisb} studied in~\cite{Morris}.
\item The `$-$' case: it corresponds to $\Lambda>0$. Then, $C_-=\sqrt{12/\Lambda}$, which is positive, thereby having a well-defined mass function given by
\be
b_-^{-\frac{1}{2}}(r)=\sqrt{\frac{12}{\Lambda}}\frac{\sqrt{\lambda}}{\left(1-\lambda\,r^2\right)},
\qquad
r< 1/\sqrt{\lambda}.
\ee
\end{itemize}

At this point, it seems natural to consider the holomorphic function given by
\be
f\left(\zeta\right)=A\,\zeta^n,
\qquad\qquad n\in {\mathbb Z^+}, 
\ee
and to explore the different PDM systems emerging from it. Note that the class of holomorphic functions that can be used is strongly restricted if we impose the conformal factor to be a radial function (recall that $f$ and $f^{-1}$ give rise to the same solution~\cite{Gibbons}). It is readily proven that in this case the conformal factor arising from~\eqref{diff} reads
\be
e^{\psi(r)}=b_\pm^{-1}(r)=C_\pm^2\frac{n^2\,{\lambda}\,r^{2n-2}}{\left(1\pm\lambda\,r^{2\,n}\right)^2}\,,
\label{ncase}
\ee
and the mass functions are just proportional to $b_\pm^{-1/2}$. 

If we compute the Gaussian curvature $K$ for the 2D metric with conformal factor~\eqref{ncase}, we obtain the  same result {\em for any value of n}, namely:
\be
K_+=\frac{4}{C^2}=-\frac{\Lambda}{3}>0,
\qquad
K_-=-\frac{4}{C^2}=-\frac{\Lambda}{3}<0,
\ee
This means that all the $n$-dependent conformal factors give rise to 2D surfaces with the same constant Gaussian curvature, regardless of $n$ and $\lambda$ (see~\cite{Laug}, p.71).
Then, as above, we have two infinite families of PDM models: 
\begin{itemize}
\item The `$+$' case, with $\Lambda<0$ and mass function given by
\be
b_+^{-\frac{1}{2}}(r)=\sqrt{\frac{12}{(-\Lambda)}}\frac{n\,\sqrt{\lambda}\,r^{n-1}}{\left(1+\lambda\,r^{2\,n}\right)}\,.
\ee

\item The `$-$' case, with $\Lambda>0$ and mass function 
\be
b_-^{-\frac{1}{2}}(r)=\sqrt{\frac{12}{\Lambda}}\frac{n\,\sqrt{\lambda}\,r^{n-1}}{\left(1-\lambda\,r^{2\,n}\right)}\,,
\qquad
r^n< 1/\sqrt{\lambda}\,.
\ee

\end{itemize}

Therefore, the size of the extra dimension for this class of  solutions is of the form
\bea
&& b_+(r)=\frac{- \Lambda}{12}
\frac{\left(1+\lambda\,r^{2\,n}\right)^2}{n^2\,\lambda\,r^{2n-2}},
\label{bdarbouxp}\\
&& b_-(r)=\frac{\Lambda}{12}
\frac{\left(1-\lambda\,r^{2\,n}\right)^2}{n^2\,\lambda\,r^{2n-2}},
\qquad
r^n< 1/\sqrt{\lambda}\,.
\label{bdarbouxm}
\eea

The mass function $b_+^{-1/2}$ is plotted in Figure 1 for $n=1,2,5$ and $\sqrt{\lambda}=1$. The same plots for $b_-^{-1/2}$ are presented in Figure 2. Figures 3 and 4 contain the plots for the estimated size $b_\pm$ of the compactified dimension, which obviously has a reciprocal behaviour when compared to the mass function. As can easily be appreciated,  the `$+$' and `$-$' models are quite different ones: the size of the extra dimension in the `$+$' case diverges for large $r$, while in the `$-$' case it vanishes in the limit $\sqrt{\lambda}\,r\to 1$.

%\bigskip

\begin{figure}[!htb]
    \centering
    \begin{minipage}{.45\textwidth}
        \centering
        \includegraphics[width=1\linewidth, height=0.25\textheight]{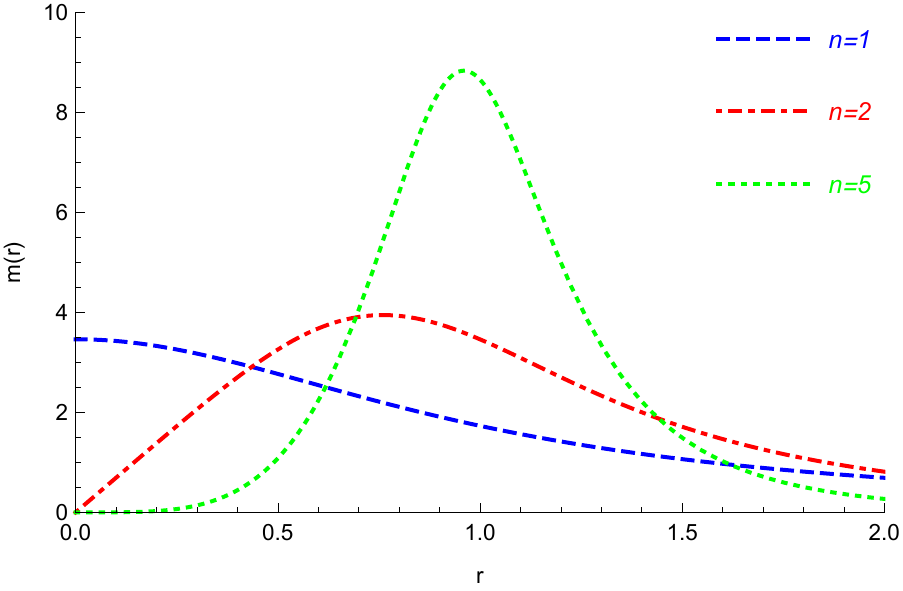}
        \caption{Mass function in the case of negative cosmological constant $\Lambda=-\left| \Lambda \right| $.}
        \label{fig:m_Case_+}
    \end{minipage}%
    \hspace{1cm}
    \begin{minipage}{0.45\textwidth}
        \centering
        \includegraphics[width=1\linewidth, height=0.25\textheight]{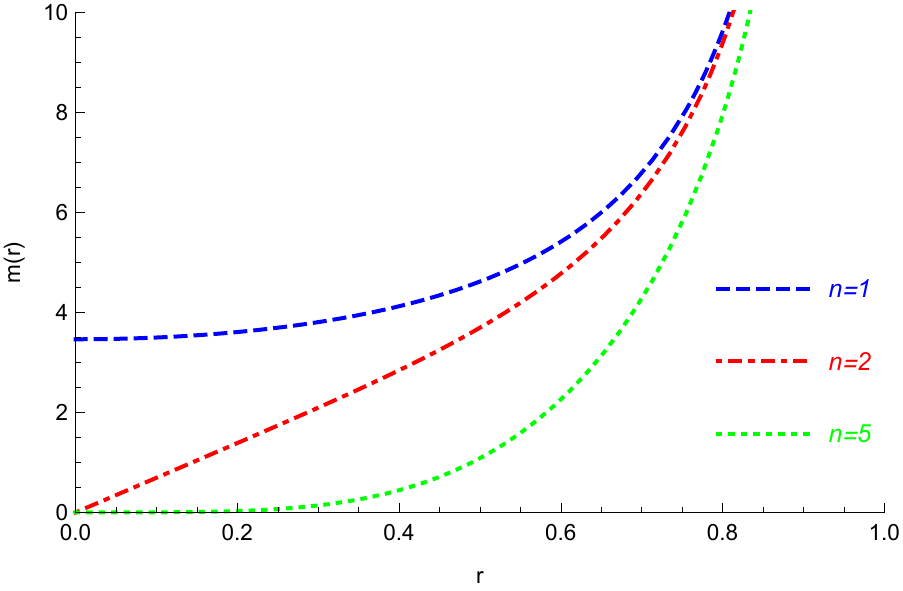}
        \caption{Mass function in the case of positive cosmological constant $\Lambda=\left| \Lambda \right| $.}
        \label{fig:m_Case_-}
    \end{minipage}
\end{figure}

%\bigskip

\begin{figure}[!htb]
    \centering
    \begin{minipage}{.45\textwidth}
        \centering
        \includegraphics[width=1\linewidth, height=0.25\textheight]{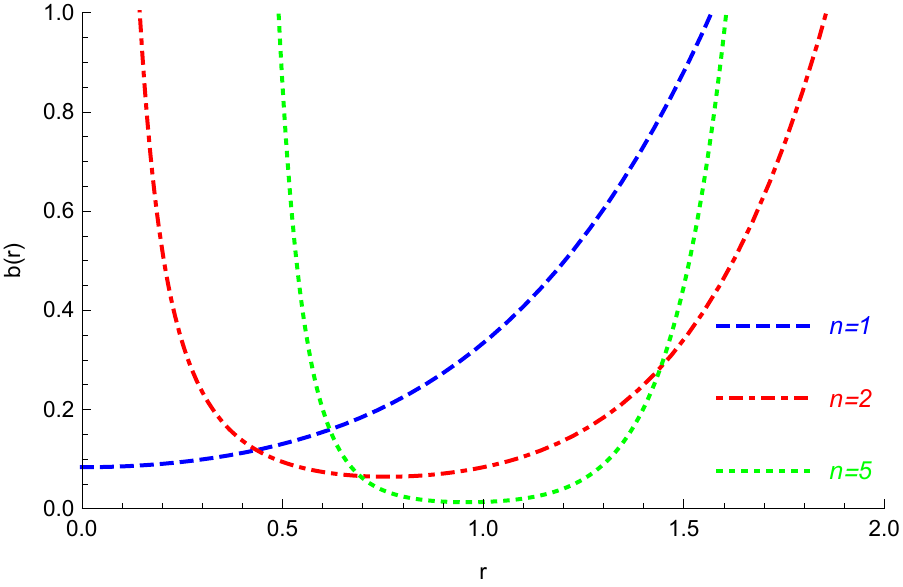}
        \caption{Size of the extra dimension in the case of negative cosmological constant $\Lambda=-\left| \Lambda \right| $.}
        \label{fig:b_Case_+}
    \end{minipage}%
        \hspace{1cm}
    \begin{minipage}{.45\textwidth}
        \centering
        \includegraphics[width=1\linewidth, height=0.25\textheight]{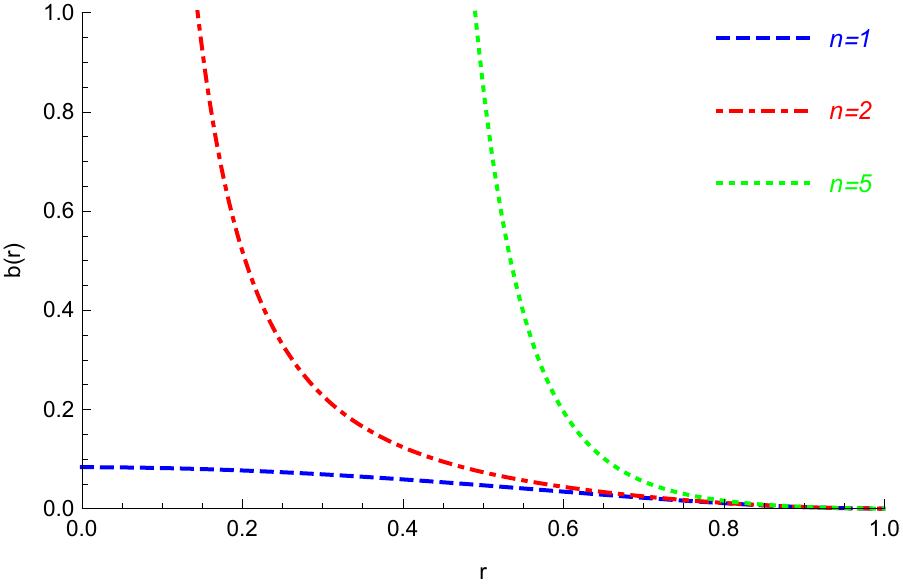}
        \caption{Size of the extra dimension in the case of positive cosmological constant $\Lambda=\left| \Lambda \right| $.}
        \label{fig:b_Case_-}
    \end{minipage}
\end{figure}

%\bigskip

%%%%%%%%%%%%%%%%%%%%%%%%%%%%%%%%%%%%%%%%%%%%%%%%%%%%%%%%%

\sect{Curvature from Kaluza-Klein compactification}

An alternative (and, to the best of our knowledge, novel) geometric interpretation of the results presented both in~\cite{Morris} and in this paper can be envisaged if we realise that the kinetic energy of the PDM Hamiltonian~\eqref{uno} can be interpreted as the one for the geodesic motion of a particle with constant mass $m_0$ on the 3D curved space defined by the (conformally flat) metric $\dd s^2= {\cal M}(\bq)\,\dd \bq^2$.
It is worth stressing that, from this perspective, the effect of the inhomogeneous compactification of the extra dimension with size $b(r)$  could be interpreted as the emergence of  a `dynamical' curvature on the 3D space originated by the variable mass function $b^{-1/2}(r)$. 

In order to be more explicit, let us recall that if we consider an $N$-dimensional conformally flat and spherically symmetric metric of the type
\begin{equation}\label{metric2}
\dd s^2=f(|\bq|)^2\,\dd\bq^2 ,
\end{equation}
where $r=|\bq|=\sqrt{\bq^2}$,  $\dd\bq^2 =\sum_{i}\dd q_i^2$ and $f$ is an
arbitrary smooth function, then the scalar curvature $R$ of the metric  is given by (see~\cite{PLB})
\begin{equation}\label{curv}
R=-(N-1)\left( \frac{    (N-4)f'(|\bq|)^2+  f(|\bq|)  \left(    2f''(|\bq|)+2(N-1)|\bq|^{-1}f'(|\bq|)  \right)}   {f(|\bq|)^4  } \right) .
\end{equation}

Therefore, since the mass function is given by $b^{-1/2}(r)$,  the kinetic energy of the anharmonic oscillator Hamiltonian characterised by~\eqref{higgs} could also be interpreted as the geodesic motion of a  particle with constant mass on the space defined by the metric
\be
 \dd s^2= \,C_\pm\frac{\sqrt{\lambda}}{(1+\la\,r^2)}\dd \bq^2 .
 \label{sosc}
 \ee
From~\eqref{curv} and by taking $N=2$, we obtain that such a space has a non-constant scalar curvature 
\be
  R=\frac{1}{C_\pm}\frac{4\,\sqrt{\lambda}}{1 + \la\, r^2},
  \label{hig}
\ee
which is twice its Gaussian curvature $K$.
As a result, the compactification generated by an extra dimension with size $b(r)\sim (1+\la\,r^2)^2$ is dynamically equivalent to the motion of a particle with constant mass on the latter curved space. Note that the space~\eqref{sosc} with $\lambda>0$ is asymptotically flat in the limit $r\to\infty$ and, when $r\to 0$, the space  has constant positive curvature.

Likewise, the corresponding geometric interpretation can be developed for the two families of compactifications that have been introduced in~\eqref{ncase}. Since the metrics induced by the mass functions would be
\be
 \dd s_\pm^2=C_\pm\frac{n\,\sqrt{\lambda}\,r^{n-1}}{\left(1\pm \lambda\,r^{2\,n}\right)}\,\dd \bq^2 ,
 \ee
the Gaussian curvature~\eqref{ccc} can straightforwardly be computed and reads
\be
  K_\pm=\frac{1}{C_\pm}\frac{2\,n\,\sqrt{\lambda}\,r^{n-1}}{(1 \pm {\lambda}\, r^{2n})}.
  \label{hig}
\ee

Figure 5 contains the corresponding plots for the Gaussian curvature of the `$+$' metric as a function of the radial coordinate $r$ for $n=1,2,5$ and $\sqrt{\lambda}=1$. The nonlinear oscillator from~\cite{Morris} corresponds to the $n=1$ case, and this turns out to be the only model with nonvanishing curvature at the origin (note that for $n=1$ the behaviour of Figures 1 and 3  is also different). As far as the $r\to \infty $ limit is concerned,  all the spaces are asymptotically flat, and the curvature is always positive for any $r$. 

The Gaussian curvature for the `$-$' metric is plotted in Figure 6. For $r=1/\sqrt{\lambda}=1$ the curvature diverges. All these spaces have positive Gaussian curvature, the $n=1$ case being again the only one featuring non-vanishing $K$ at the origin. Again, Figures 2 and 4 show the corresponding features of these models in terms of the mass and the size functions.

%\bigskip

\begin{figure}[!htb]
    \centering
    \begin{minipage}{.45\textwidth}
        \centering
        \includegraphics[width=1\linewidth, height=0.25\textheight]{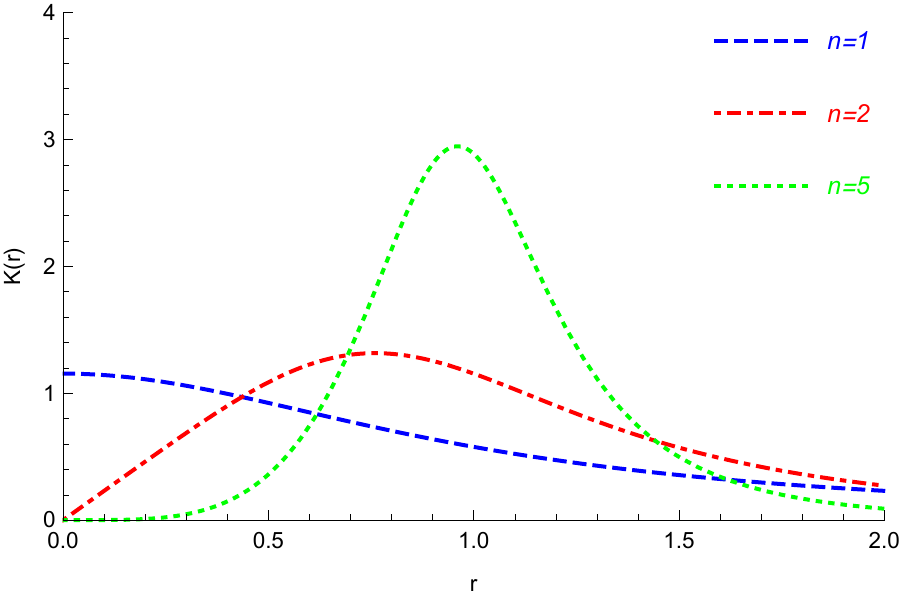}
        \caption{Gaussian curvature in the case of negative cosmological constant $\Lambda=-\left| \Lambda \right| $.}
        \label{fig:b_Case_+}
    \end{minipage}%
        \hspace{1cm}
    \begin{minipage}{.45\textwidth}
        \centering
        \includegraphics[width=1\linewidth, height=0.25\textheight]{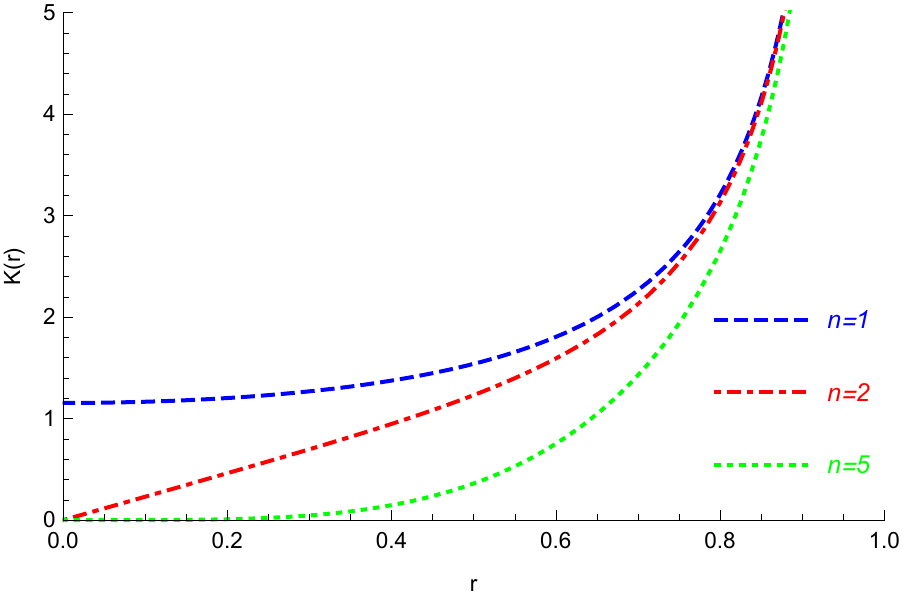}
        \caption{Gaussian curvature in the case of positive cosmological constant $\Lambda=\left| \Lambda \right| $.}
        \label{fig:b_Case_-}
    \end{minipage}
\end{figure}

%%%%%%%%%%%%%%%%%%%%%%%%%%%%%%%%%%%%%

\section{Concluding remarks}

Both the role of PDM quantum systems in condensed matter models and the integrability properties of PDM Hamiltonians in classical and quantum contexts are well-known, as it is demonstrated in the above-mentioned bibliography. Nevertheless, the generation of PDM systems as a result of compactification processes of an extra dimension seems to be a novel feature that, in our opinion, deserves further investigation. Indeed, the Kaluza-Klein nature of the latter construction suggests that a deeper insight into the geometric properties of PDM systems could be pertinent.

In this paper we have faced this issue from two different viewpoints. On one side, we have provided two (quite different) infinite families of PDM systems that can be obtained as the result of the compactification of a dilatonic field, thus generalising the result presented in~\cite{Morris} and providing many new examples. On the other side, the geometric features of all these systems have been explored in two directions. Firstly, by showing that the square of the mass function characterising all of them (including the one presented in~\cite{Morris}) is just the conformal factor for a set of two-dimensional metrics with constant Gaussian curvature. Secondly, by interpreting the appearance of the variable mass as a geometric effect that generates curvature as a dynamical consequence of the compactification of the extra dimension on each point of the Euclidean manifold. 

Indeed, several issues on this subject remain open, and their study would be valuable in order to elucidate the physical and geometric content of the Kaluza-Klein compactification framework here used. In particular, it would be interesting to study the integrability properties and the explicit solutions for the quantum motion of a particle on the two new families of spaces (or, equivalently, of the two families of PDM quantum Hamiltonians) here introduced, a task that could be performed by following the methods employed in~\cite{CRS04, CRS07, SMjpa, darboux, darbouxPLA}.  Also, PDM systems without radial symmetry  arising from~\eqref{dilaton} can be also constructed and analysed from this geometrical viewpoint. Finally, we recall that the so-called quantum deformations of the underlying space-time symmetries have been also found to be responsible of the emergence of curvature~(see \cite{BHRplb}), and spin interacting models can be also thought of as a source of position dependent mass and of effectively curved backgrounds (see~\cite{Der1,Der2}). Therefore, possible relationships between the latter approaches and the results here presented could be explored.
Work on all these lines is in progress and will be presented elsewhere.

 \section*{{Acknowledgements}}

This work was partially supported by the Spanish Ministerio de Econom\1a y Competitividad    (MINECO) under project MTM2013-43820-P and by Junta de Castilla y Le\'on  under grants  BU278U14 and VA057U16. I.G-S. acknowledges a predoctoral grant from Junta de Castilla y Le\'on and the European Social Fund. The authors are indebted to F.J. Herranz and M. Santander for useful discussions, and to J.R. Morris for relevant remarks on a previous version of this work.

%%%%%%%%%%%%%%%%%%%%%%%%%%%%%%%%%%%%%%%%%%%%%%

%%%%%%%%%%%%%%%%%%%%%%%%%%%%%%

\end{document}